\documentstyle[twocolumn,graphics,prl,aps,floats]{revtex}
\begin{document}
\draft 
\wideabs{ 
\title{Heat Transport in a Strongly Overdoped Cuprate: \\
Fermi Liquid and Pure $d$-wave BCS Superconductor}

\author{Cyril Proust$^{1\dagger}$, Etienne Boaknin$^1$, R.W. Hill$^1$,
Louis Taillefer$^1$, and A.P. Mackenzie$^2$}

\address{ $^1$ Canadian Institute for Advanced Research, Department of
Physics, University of Toronto, Toronto, Ontario, Canada}


\address{$^2$ School of Physics and Astronomy,
University of St. Andrews, St. Andrews, Fife KY16 9DE, Scotland}

\date{\today}

\maketitle

\begin{abstract}

  The transport of heat and charge in the overdoped cuprate
  superconductor Tl$_2$Ba$_2$CuO$_{6+\delta}$ was measured down to low
  temperature. In the normal state, obtained by applying a magnetic
  field greater than the upper critical field, the Wiedemann-Franz law
  is verified to hold perfectly. In the superconducting state, a large
  residual linear term is observed in the thermal conductivity, in
  quantitative agreement with BCS theory for a $d$-wave
  superconductor. This is compelling evidence that the electrons in
  overdoped cuprates form a Fermi liquid, with no indication of
  spin-charge separation.

\end{abstract}
 }

A fundamental question about the rich and baffling behavior of
electrons in high-temperature superconductors is whether or not it can
be understood in the framework of Fermi-liquid (FL) theory, the
standard theory of electrons in solids. Several authors believe that
when the concentration of electronic carriers in these cuprate
materials is sufficiently low, as in the so-called underdoped region
of the doping phase diagram, the basic excitations of the electron
system are not the usual Landau quasiparticles characteristic of FL
theory. In one class of proposals \cite{Kivelson,Anderson,Fisher} for
example, the electron is thought to fractionalize into a neutral
spin-carrying excitation, called a ``spinon'', and a spinless
charge-carrying excitation, called a ``holon'' or ``chargon''.  However to
this day, such ``spin-charge separation'' has not been confirmed
experimentally. On the other hand, after 15 years of intensive research it is
still not known with any certainty whether or not the ground state of
cuprates is a Fermi-liquid in any region of the phase diagram.
It is widely assumed that in the metallic-like overdoped regime at
high carrier concentration FL theory does hold, but there is little
solid evidence to support this lore. 

In this Letter, we present the results of a study which show that
strongly overdoped cuprates do not undergo spin-charge separation and
their ground state is most likely a Fermi liquid.  By measuring the
transport of both heat and charge in the normal state at very low
temperature, we were able to verify that
one hole-doped cuprate in the overdoped regime obeys the
Wiedemann-Franz 
(WF) law.  This universal law is a robust signature of FL
theory, stating simply that the electronic carriers of heat are
fermionic excitations of charge $e$.  In addition, the thermal
conductivity in the superconducting state is found to be in 
good
agreement with BCS theory for a superconductor with a pure $d$-wave
order parameter.

The particular compound chosen for this study is
Tl$_2$Ba$_2$CuO$_{6+\delta}$, because it can easily be overdoped.  In
many ways, it is the ideal cuprate material. Its crystal structure is
tetragonal, without the CuO chains that complicate the properties of
the orthorhombic compounds YBa$_2$Cu$_3$O$_{7-\delta}$ (Y-123) and
YBa$_2$Cu$_4$O$_8$ (Y-124), or the buckling that alters the unit cell
of Bi$_2$Sr$_2$CaCu$_2$O$_8$ (Bi-2212).  It is made of a stack of
single CuO$_2$ planes, and is therefore not subject to possible
bi-layer effects such as encountered in Bi-2212. It has a high maximum
critical temperature $T_c^{max}$ of 90 K, at optimal doping, much as
in Y-123 and Bi-2212. In this sense, it is free of the possible
concerns about the low $T_c$ found in single-plane
La$_{2-x}$Sr$_x$CuO$_4$ (LSCO). Finally, the $d_{x^2-y^2}$ symmetry of
its superconducting state has been confirmed by
phase-sensitive measurements, at least at optimal doping \cite{Tsuei}.

Mackenzie {\it et al.} measured the resistivity of a
strongly-overdoped crystal of Tl-2201, with $T_c =15$~K
\cite{Mackenzie}.  In zero magnetic field, $\rho(T)$ was found to
follow roughly a power law of $T^{1.8}$ from room temperature down to
$T_c$ and extrapolate to $\rho_0 \simeq 7~\mu \Omega$~cm at $T=0$.
The resistive upper critical field at $T \to 0$, $H_{c2}(0)$, is
between 12 and 16 T (for fields perpendicular to the conducting
CuO$_2$ planes, {\it i.e.}  $H \parallel c$), depending on the precise
criterion.

The overdoped samples of Tl-2201 used in this study were rectangular
single crystals with typical dimensions of $0.4$~mm and $0.2$~mm in
the tetragonal basal plane and $10~\mu$m along the $c$-axis.  The
voltage pads had a width of $25~\mu$m and the spacing between the
electrodes was $0.3$~mm. They were grown by the same technique as used
by Mackenzie and co-workers in previous studies
\cite{Mackenzie,Carrington-C,Carrington-R}. They have $T_c\simeq15$~K,
in zero magnetic field.  Using the empirical formula $T_c/T_c^{max} =
1 - 82.6 (p-0.16)^2$, this translates into a carrier concentration of
$p = 0.26$~hole/Cu atom. To obtain such critical temperatures, the
samples were annealed in 1 bar of flowing O$_2$ at 350 $^o$C for two
days. The resistivity of our samples is essentially identical to that
obtained previously \cite{Mackenzie}, with $\rho_0 = 5.6~\mu
\Omega$~cm. Both heat and charge transport were measured using the
same contacts, made by diffusing silver epoxy.  A typical value for
the contact resistance was 0.1 $\Omega$ at 4~K. The thermal
conductivity was measured down to below 100 mK 
with a standard one-heater two-thermometer technique in a dilution
refrigerator. The magnetic field was
applied along the $c$-axis. 
The geometric factor used to convert from 
resistance (electrical
or thermal) to electrical resistivity $\rho$ or thermal conductivity
$\kappa$ was set by requiring that $\rho(300~{\rm K}) = 180~\mu
\Omega$~cm, the value obtained in previous studies of numerous 
crystals with the same doping level
\cite{Mackenzie,Carrington-C,Carrington-R,Mackenzie2}. The uncertainty
on this value is estimated at $\pm 10~\mu \Omega$~cm.

The resistivity is shown 
in Fig.~1, for
fields ranging from zero to above $H_{c2}(0)$.  A slight positive 
magnetoresistance is observed, in
agreement with previous work \cite{Carrington-R}.  The resistivity
below 30~K (and above $T_c$) is best fit by the
function $\rho = \rho_0 + bT + cT^2$, with a substantial linear term
({\it i.e.} $bT > cT^2$ for $T < 15$~K). The fitting parameters are
$\rho_0(H) = 5.84$, 5.99 and 6.15 $\mu \Omega$~cm at $H = 7$, 10 and
13 T, respectively, and $b=0.064~\mu \Omega$~cm~K$^{-1}$,
$c=0.0054~\mu \Omega$~cm~K$^{-2}$ at 13 T. This unusual dependence was
reported previously \cite{Mackenzie2} and interpreted as
``non-Fermi-liquid'' behavior, in the sense that no linear term is
expected in conventional FL theory. Deviations from the standard $T^2$
dependence have been observed in a number of heavy-fermion materials,
for example, $T^{1.2}$ in CePd$_2$Si$_2$ below 20~K \cite{NFL}. In
these materials, this is associated with the proximity to a quantum
critical point (QCP), where antiferromagnetic order sets in as a
function of pressure or chemical composition. In the case of cuprates,
the obvious QCP would be the onset of superconductivity at a critical
concentration $p_c$ close to 0.3 hole/Cu atom, but a QCP has also been
postulated to exist inside the
superconducting region.
\begin{figure}
\centering
\resizebox{\columnwidth}{!}{\includegraphics*{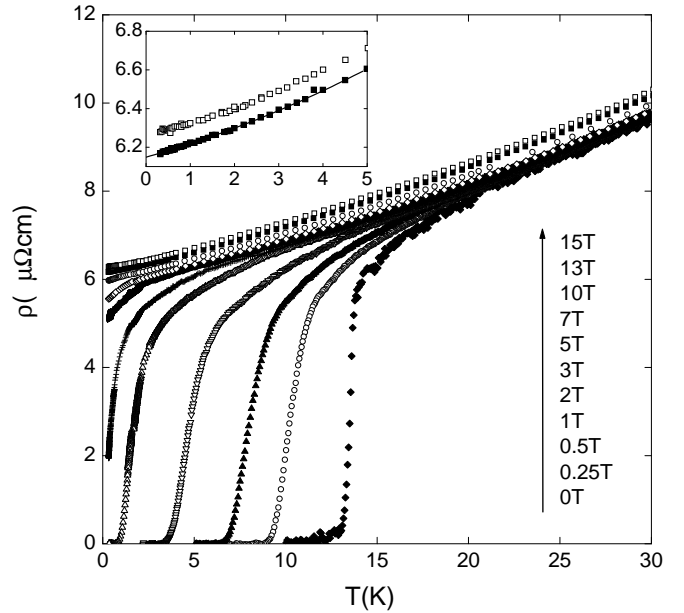}}
\vspace{4pt}
\caption{Electrical resistivity of Tl-2201 vs temperature for a
current in the basal plane at different values of the magnetic field
applied normal to the plane. All trace of superconductivity has
vanished by 13~T. {\it Inset}: $\rho(T)$ at $H = 13$~T (filled
symbols) and $15$~T (open symbols). The line is a fit of the $13$~T
data to the functional form $\rho(T) = \rho_0 + bT +cT^2$.}
\label{Fig1}
\end{figure}
\noindent 

The thermal conductivity $\kappa$ is shown in Fig.~2. The data is
plotted as $\kappa/T$ vs $T^2$ to separate the contribution of
electrons from that of phonons, given that the asymptotic dependence
of the former as $T \to 0$ is linear in $T$ while that of the latter
is cubic. In other words, in Fig.~2, the electronic contribution is
the residual linear term $\kappa_0/T$ given by the intercept of a
linear fit with the $T=0$ axis. The value of $\kappa_0/T$ obtained in
this way is: 
 1.41, 2.76, 3.47, 3.75, 3.87, 3.90, 3.95, and 3.95
mW~K$^{-2}$~cm$^{-1}$, at $H = 0, 1, 2.5, 4, 5.5, 7, 10,$ and $13$~T,
respectively.  As explained above, the uncertainty on the overall
absolute value is approximately $\pm 5\%$.  However, the relative
uncertainty, {\it e.g.}  between different fields, is much lower,
around 1 \%. This high degree of reliability 
is due to the fact that in these samples electrons
conduct much better than phonons, and hence the slope of $\kappa(T)/T$
in Fig.~2 is weak relative to the intercept. Note that at high fields,
electrons scatter phonons very effectively and $\kappa(T)$ is entirely
electronic below 1~K.


Fundamentally, the linear term in $\kappa$ at $T=0$ reveals the
presence of fermionic excitations in the electron system. 
We can then ask whether these excitations carry charge.
This question can only be addressed in the absence of any superfluid
that can also carry charge, which amounts to testing the
WF law in the normal state.  This law is one of
the most fundamental properties of a Fermi liquid, reflecting the fact that the
ability of a quasiparticle to transport energy is the same as its
ability to transport charge, provided it cannot lose energy through
collisions.  It states that the heat conductivity $\kappa$ and the
electrical conductivity $\sigma$ of a metal are related by a universal
constant:
\begin{equation}
\frac{\kappa}{\sigma T} = \frac{\pi^2}{3} ~ (\frac{k_B}{e})^2 \equiv L_0
\end{equation}
where $T$ is the absolute temperature, $k_B$ is Boltzmann's constant
and $L_0= 2.44 \times 10^{-8}$~W~$\Omega$~K$^{-2}$ is Sommerfeld's
value for the Lorenz ratio $L \equiv \kappa/\sigma T$
.  Theoretically, electrons are predicted to obey the WF law at $T \to
0$ in a
 wide range of environments: in both three
 or
two dimensions (but not strictly in one dimension), for any strength
of disorder and interaction \cite{Castellani}, scattering and magnetic
field \cite{Kearney}.  Experimentally, the WF law does appear to be
universal at $T \to 0$: until recently, no material had been reported
to violate it.
The first exception was found in optimally-doped
Pr$_{2-x}$Ce$_x$CuO$_4$ (PCCO), an electron-doped cuprate 
\cite{Hill}.

It is in general difficult to test the WF law in cuprate
superconductors because of their high upper critical fields.
In our crystals, the superconductivity has
completely vanished by 13 T, at which field we find $\kappa_0/T = 3.95
\pm 0.04$~mW~K$^{-2}$~cm$^{-1}$ and $\rho_0 = 6.15 \pm 0.03~\mu
\Omega$~cm, so that $L = \rho_0 \kappa_0/T = 0.99 \pm 0.01 ~L_0$, in
perfect agreement with the WF law.  Note that the Lorenz ratio does
not suffer from the 5\% uncertainty associated with the geometric
factor, as both transport measurements are performed using the same
sample with the same contacts. The error bars are therefore on the
order of 1~\%. In Fig.~2, the transport of heat and charge are
compared directly by reproducing the charge conductivity at $13$~T
from Fig.~1. This is done by plotting $L_0/\rho(T)$ vs $T$ using the
fit to the $13$~T data for $\rho(T)$ (inset of Fig.~1).
The charge conductivity $L_0\sigma(T)$ 
is seen to be equal to the heat conductivity $\kappa(T)/T$ at $13$~T.

The basic implication of this result is that {\sl the fermions which
carry heat also carry charge $e$ and are therefore indistinguishable
from standard Landau quasiparticles}.  In particular, there is no
evidence of any spin-charge separation.  Indeed, if electrons were to
fractionalize into neutral spin-carrying fermions (spinons) and
charged bosons (chargons) \cite{Fisher}, there would be no reason to
expect the 
WF law to hold, as the heat-carrying fermions would not take part in
the transport of charge.  This result therefore imposes a constraint
on theories of spin-charge separation (SCS): the critical hole
concentration $p_{SCS}$ at which electron fractionalization starts to
occur is not the QCP where superconductivity starts to occur (on the
overdoped side of the phase diagram), but can only be lower.  In other
words, any hypothetical onset of SCS must obey $p_{SCS} < 0.26 < p_c$.
{\sl It therefore appears that the mechanism for superconductivity in
this overdoped region of the phase diagram is not the condensation of
charge-$e$ bosons}, but most likely Cooper pairing.  Note that 
(barring any profound electron-hole asymmetry) this conventional picture 
is expected to break down with underdoping, as suggested by the violation 
of the WF law in PCCO near optimal doping \cite{Hill}.

Although the standard FL description fails, as revealed by the
non-quadratic $T$ dependence of $\rho(T)$, the basic nature of the
electronic excitations in the limit of zero energy is that of Landau
FL quasiparticles.  (A similar situation is seen in heavy-fermion
materials \cite{Kambe}.) 
\begin{figure}
\centering
\resizebox{\columnwidth}{!}{\includegraphics*{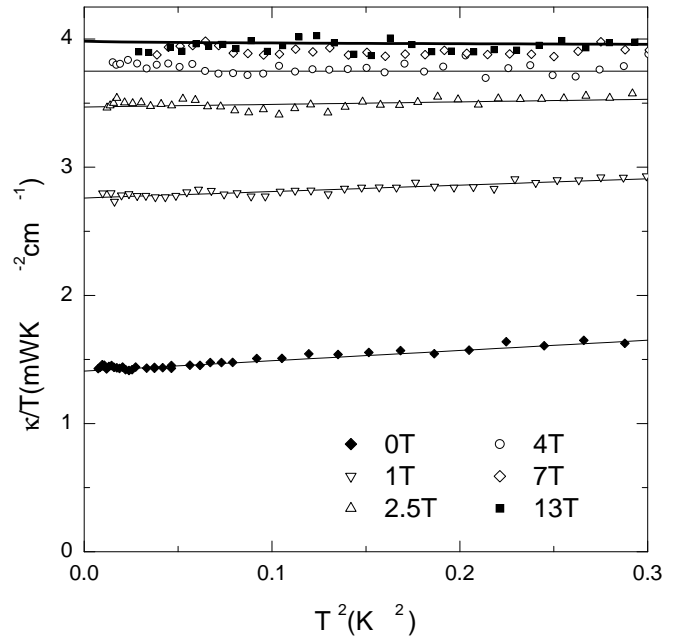}}
\vspace{4pt}
\caption{Thermal conductivity of Tl-2201 for a heat current in the
basal plane, plotted as $\kappa/T$ vs $T^2$, at different values of
the magnetic field applied normal to the plane. The thin lines are
linear fits to the data. The thick line is $L_0/\rho(T)$ where
$\rho(T)$ is a fit to the resistivity at 13~T (see inset of Fig.~1).
}
\label{Fig2}
\end{figure}

In the absence of a magnetic field, there is a large residual linear
term in the thermal conductivity of Tl-2201, namely $\kappa_0/T$ =
1.41 mW K$^{-2}$ cm$^{-1}$.  A similar term has also been observed in
other hole-doped cuprates, albeit at optimal doping, where it is much
smaller: $\kappa_0/T$ = 0.14, 0.15 and 0.11 mW K$^{-2}$ cm$^{-1}$, in
Y-123 \cite{Taillefer}
, Bi-2212 \cite{Chiao2,Behnia} and LSCO
\cite{Taillefer2}, respectively.  Within BCS theory applied to a
$d$-wave superconductor, this residual heat conduction is expected,
arising from zero-energy quasiparticles induced by impurity scattering
near the nodes in the $d_{x^2-y^2}$ gap function.  In the clean limit,
where the scattering rate $\Gamma \ll k_B T_c / \hbar$, it is
universal (in the sense that it is independent of $\Gamma$) and it
depends only on the ratio of the two quasiparticle velocities ($v_F$
and $v_2$) which govern the Dirac-like spectrum of nodal
quasiparticles, $E = \hbar \sqrt{ v_F^2 k_1^2 + v_2^2 k_2^2 }$
\cite{Durst}:
\begin{equation}
 \frac{\kappa_0}{T} = \frac{k_B^2}{3 \hbar} \frac{n}{c} ( \frac{v_F}{v_2} + \frac{v_2}{v_F} )
\end{equation}
where $n$ is the number of CuO$_2$ planes per unit cell of height $c$
(along the $c$-axis), and $\vec{k_1}$ and $\vec{k_2}$ are unit vectors
pointing in directions normal and tangential to the Fermi surface at
the node, respectively. In other words, $v_F$ is the Fermi velocity in
the nodal direction and $v_2$ is proportional to the slope of the gap
at the node, $d \Delta / d \phi = \hbar k_F v_2$, with $k_F$ the Fermi
wavevector.

Applying Eq.~2 to Tl-2201, for which $n=2$ and $c=23.2~\AA$, we get
$v_F/v_2 = 270$.  A rough estimate using Fermi surface parameters
typical of cuprates, namely $v_F = 2.5 \times 10^7$~cm/s and $k_F =
0.7~\AA^{-1}$ (the values measured in Bi-2212 \cite{Mesot}), and the
simplest $d$-wave gap function, $\Delta = \Delta_0 {\rm cos}2\phi$,
with the weak-coupling relation for a $d$-wave superconductor,
$\Delta_0 = 2.14~k_B T_c$, gives $v_F/v_2 = 210$. This shows that the
magnitude of $\kappa_0/T$ is in good agreement with the simplest BCS
analysis.

It should be recognized that even though the mean free path in these
samples is rather long (in the range $500-1000~\AA$
\cite{Mackenzie2}), the scattering rate $\Gamma$ is not small compared
to $T_c$. It may be estimated using the standard 
expression
for the normal state conductivity: $\kappa_N/T = \frac{1}{3} \gamma_N
v_F^2 \tau$, where $\gamma_N$ is the specific heat coefficient and
$\tau= 1 / (2\Gamma)$.  With $\gamma_N \simeq 3$~mJ~K$^{-2}$
mole$^{-1}$ \cite{Carrington-C} and $v_F = 2.5 \times 10^7$~cm/s, one
gets $\hbar \Gamma \simeq 0.4~k_B T_c$. At finite $\Gamma$,
corrections to Eq.~2 give an increase in $\kappa_0/T$
\cite{Maki}. Assuming $\Delta_0=2.14~k_B T_c$, the correction for
$\hbar \Gamma / k_B T_c = 0.4$ is by a factor of approximately $1.5$
\cite{Maki}. 
Thus the correct value of $v_F/v_2$ is
probably closer to $270/1.5 = 180$ \cite{Note1}.

It will be interesting to investigate the doping dependence of
$\kappa_0/T$ as a way of measuring the dependence of the gap function
on carrier concentration, via $v_2$. In the absence of further data on
Tl-2201, we may compare with optimally-doped Y-123 ($T_c = 93$~K) or
Bi-2212 ($T_c \simeq 90$~K), for which $v_F/v_2 = 14$ and 19,
respectively \cite{Chiao2}. (The value of 19 for Bi-2212 agrees very
well with the value of 20 obtained from ARPES measurements of $v_F$
and $v_2$ separately \cite{Chiao2,Mesot}.)  Under the assumption of a
doping independent $v_F$, verified in both Bi-2212 \cite{Mesot} and
LSCO \cite{Lanzara}, one immediately sees that $v_2$, or the magnitude
of the gap (near the nodes), scales roughly with $T_c$.  This strongly
suggests that the standard BCS relation between gap magnitude and
transition temperature, $\Delta_0 \propto T_c$, holds in the overdoped
regime.  This is in striking contrast with what is found in the
underdoped region of the phase diagram.  Indeed, 
our
measurements on underdoped Y-123 and LSCO \cite{Hawthorn} reveal that
$v_F/v_2$ {\it decreases} as $T_c$ is reduced by underdoping (see also
\cite{Ando}).

Several authors have proposed the existence of a 
QCP within the superconducting dome in the phase diagram of
cuprates, either as a theoretical prediction to explain the diagram
itself or as suggested in various experiments.  Its location is
usually taken to be near 
optimal doping, in the
neighbourhood of $p = 0.2$. If it is associated with a change in the
symmetry of the superconducting order parameter, Vojta {\it et al.}
have argued that the most likely scenario is a transition from a pure
$d_{x^2-y^2}$ state to a complex order parameter of the form
$d_{x^2-y^2} + ix$, where $x$ can have either $s$ or $d_{xy}$ symmetry
\cite{Sachdev}. Dagan and Deutscher have recently reported 
a split
zero-bias anomaly in their tunneling on Y-123 thin films as soon as
the material is doped beyond optimal doping, a feature which they
attribute to the appearance of a complex component to the order
parameter in the bulk 
\cite{Deutscher}. The presence of a subdominant
component $ix$ in the order parameter causes the nodes to be removed,
as the gap can no longer go to zero in any direction.  The observation
of a residual linear term in the thermal conductivity, a direct
consequence of nodes in the gap, 
therefore excludes the possibility of any such
subdominant order parameter.  (More precisely, since our measurement
goes down to 
100~mK, it puts an upper bound on the magnitude of
$|x|$ relative to $|d_{x^2-y^2}|$ at about 0.5~\%.) Moreover, there is
no subdominant order parameter in Tl-2201 at optimal doping 
\cite{Tsuei}.
In other words, {\sl if
there truly is a QCP between optimal doping at $p \simeq 0.16$ and the
critical point $p_c \simeq 0.3$, it does not appear to be associated
with the onset of a complex component in the order parameter}.

In summary,
the low-temperature transport properties of Tl-2201 with $T_c = 15$~K
show that spin-charge separation does not occur in strongly overdoped
cuprates.  The normal state at $T \to 0$ satisfies the Wiedemann-Franz
law perfectly, demonstrating that the only electronic excitations
carrying heat and charge are Landau quasiparticles.  The
superconducting state obeys BCS theory in that the residual heat
conduction is 
of the expected magnitude for a 
$d$-wave
gap and the dependence of the low-energy spectrum on doping strongly
suggests that the gap scales with $T_c$ in the conventional way.
Finally, the possibility of a sub-dominant order parameter ($ix$) is
ruled out.



We are grateful to D. Hawthorn, C. Lupien, J. Paglione, F. Ronning,
and M. Sutherland for help in various aspects of the measurements, and
to S. Kivelson, C. Lannert and K. Behnia for useful discussions. This
work was supported by the Canadian Institute for Advanced Reasearch
and funded by NSERC of Canada.\\
$\dagger$ Present address: Laboratoire National des Champs
Magn\'etiques Puls\'es, 
31432 Toulouse, France.


\end{document}